\documentclass{llncs}

\usepackage{epsfig}
\usepackage{array}

\begin{document}

\title{Droplet Spreading on Heterogeneous Surfaces \\
  using a Three-Dimensional \\
  Lattice Boltzmann Model}

\titlerunning{Spreading on Heterogeneous Surfaces using a 3D LB Model}

\author{A. Dupuis\thanks{Contact: \email{dupuis@thphys.ox.ac.uk}}
  \and A.J. Briant \and C.M. Pooley  \and J.M. Yeomans}

\authorrunning{A.Dupuis et al.}

\institute{Department of Physics, Theoretical Physics, University of Oxford,\\
  1 Keble Road, Oxford OX1 3NP, UK.}

\newcommand{\pos}{\ensuremath{\mathbf{r}}}
\newcommand{\dt}{\ensuremath{\Delta t}}
\newcommand{\dr}{\ensuremath{\Delta \pos}}
\newcommand{\vi}{\ensuremath{\mathbf{v}_i}}
\newcommand{\vm}{\ensuremath{v}}
\renewcommand{\u}{\ensuremath{\mathbf{u}}}
\newcommand{\dab}{\ensuremath{\delta_{\alpha\beta}}}
\newcommand{\eq}[1]{equation (\ref{#1})}
\newcommand{\Eq}[1]{Equation (\ref{#1})}
\newcommand{\fig}[1]{fig. \ref{#1}}
\newcommand{\Fig}[1]{Fig. \ref{#1}}

\maketitle

\begin{abstract}
  We use a three-dimensional lattice Boltzmann model to investigate
  the spreading of mesoscale droplets on homogeneous and heterogeneous
  surfaces. On a homogeneous substrate the base radius of the droplet
  grows with time as $t^{0.28}$ for a range of viscosities and surface
  tensions. The time evolutions collapse onto a single curve as a
  function of a dimensionless time. On a surface comprising of
  alternate hydrophobic and hydrophilic stripes the wetting velocity
  is anisotropic and the equilibrium shape of the droplet reflects the
  wetting properties of the underlying substrate.
\end{abstract}

\section{Introduction}
\label{sec:intro}

Wetting processes, such as the spreading of a droplet over a surface,
have attracted the attention of scientists for a long
time~\cite{degennes:85}. A great deal is understood about the wetting
behaviour of equilibrium droplets. However less is known about the
dynamics of these systems, a problem of considerable industrial
relevance with the advent of ink-jet printing. The droplets involved
in printing typically have length scales of microns. Experimental work
on such mesoscopic droplets is difficult and expensive because of the
length- and time-scales involved. Therefore there is a need for
numerical modelling both to investigate the physics and to help design
and interpret the experiments.

Lattice Boltzmann models are a class of numerical techniques ideally
suited to probing the behaviour of fluids on mesoscopic length
scales~\cite{succi-book:01}. Several lattice Boltzmann algorithms for
a liquid-gas system have been reported in the
literature~\cite{swift:96,shan:93,he:98}. They solve the Navier-Stokes
equations of fluid flow but also input thermodynamic information,
typically either as a free energy or as effective microscopic
interactions. They have proved successful in modelling such diverse
problems as fluid flows in complex geometries~\cite{succi:89},
two-phase models~\cite{swift:96,shan:93}, hydrodynamic phase
ordering~\cite{kendon:99} and sediment transport in a
fluid~\cite{dupuis:00b}.

Here we show that it is possible to use a lattice Boltzmann approach
to model the spreading of mesoscale droplets and, in particular, to
illustrate how a droplet spreads on a substrate comprising of
hydrophilic and hydrophobic stripes. 


We consider a one-component, two-phase fluid and use the free energy
model originally described by Swift et al.~\cite{swift:96} with a
correction to ensure Galilean invariance~\cite{holdych:98}. The
advantage of this approach for the wetting problem is that it allows
us to tune equilibrium thermodynamic properties such as the surface
tension or static contact angle to agree with analytic predictions.
Thus it is rather easy to control the wetting properties of the
substrate. Three dimensional simulations of spreading on smooth and
rough substrates have previously been reported in~\cite{raiskinmaki:00}
using a different lattice Boltzmann algorithm.

The paper is organised as follows. First we summarise the main
features of the lattice Boltzmann approach. The model is validated by
showing the consistency of the measured equilibrium contact angle with
that predicted by Young's law and by measuring the base radius of the
spreading droplet as a function of time obtaining, as expected, a
power law growth. We show that when the reduced base radius is plotted
as a function of reduced time the data fall on a universal curve for
several values of surface tension and viscosity.

We then consider spreading on a heterogeneous substrate consisting of
alternate hydrophobic and hydrophilic stripes. We find that the
spreading velocity is anisotropic and that the final droplet shape
reflects the wetting properties of the underlying substrate. Finally,
a conclusion suggests extensions to the work presented here.

\section{Simulating spreading}
\label{sec:model}

\subsection{The lattice Boltzmann model}

The lattice Boltzmann approach solves the Navier-Stokes equations by
following the evolution of partial distribution functions $f_i$ on a
regular, $d$-dimensional lattice formed of sites $\pos$. The label $i$
denotes velocity directions and runs between $0$ and $z$. $DdQz+1$ is
a standard lattice topology classification. The $D3Q15$ lattice
topology we use here has the following velocity vectors $\vi$:
$(0,0,0)$, $(\pm 1,\pm 1,\pm 1)$, $(\pm 1,0,0)$, $(0,\pm 1, 0)$, $(0,
0, \pm 1)$ in lattice units.

The lattice Boltzmann dynamics are given by
\begin{equation}
f_i(\pos+ \dt \vi,t+\dt)=f_i(\pos,t)
                        +\frac{1}{\tau}\left(f_i^{eq}(\pos,t)-f_i(\pos,t)\right)
\label{eq:lbDynamics}
\end{equation}
where $\dt$ is the time step of the simulation, $\tau$ the relaxation
time and $f_i^{eq}$ the equilibrium distribution function which is a
function of the density $n=\sum_{i=0}^z f_i$ and the fluid velocity
$\u$ defined through the relation $n\u=\sum_{i=0}^z f_i\vi$.

The relaxation time tunes the kinematic viscosity as
\begin{equation}
\nu=\frac{\dr^2}{\dt} \frac{C_4}{C_2} (\tau-\frac{1}{2})
\label{eq:visco}
\end{equation}
where $\dr$ is the lattice spacing and $C_2$ and $C_4$ are coefficients
related to the topology of the lattice. These are equal to $3$ and $1$
respectively when one considers a $D3Q15$ lattice
(see~\cite{dupuis:02} for more details).

It can be shown~\cite{swift:96} that equation~(\ref{eq:lbDynamics})
reproduces the Navier-Stokes equations of a non-ideal gas if the local
equilibrium functions are chosen as
\begin{eqnarray}
f_i^{eq}&=&A_\sigma+B_\sigma u_\alpha v_{i\alpha} + C_\sigma \u^2
         +D_\sigma u_\alpha u_\beta v_{i\alpha}v_{i\beta} 
         +G_{\sigma\alpha\beta} v_{i\alpha}v_{i\beta}, \quad i>0,
         \nonumber \\
f_0^{eq}&=& n - \sum_{i=1}^z f_i^{eq}
\label{eq:lbEq}
\end{eqnarray}
where Einstein notation is understood for the Cartesian labels
$\alpha$ and $\beta$ (i.e.  $v_{i\alpha}u_\alpha=\sum_\alpha
v_{i\alpha}u_\alpha$) and where $\sigma$ labels velocities of
different magnitude.

The coefficients $A_\sigma$, $B_\sigma$, $C_\sigma$, $D_\sigma$ and
$G_\sigma$ are chosen so as to satisfy the relations
\begin{eqnarray}
\sum_i f_i^{eq} & = & n, \nonumber \\
\sum_i f_i^{eq} v_{i\alpha} & = & n u_{\alpha},  \nonumber \\
\sum_i f_i^{eq} v_{i\alpha} v_{i\beta} & = & P_{\alpha\beta} + n
u_\alpha u_\beta + \nu \left( u_\alpha \partial_\beta n + u_\beta
  \partial_\alpha n + u_\gamma \partial_\gamma n \dab \right),  \nonumber \\
\sum_i f_i^{eq} v_{i\alpha} v_{i\beta} v_{i\gamma} & = & \frac{c^2 n}{3}
(u_\alpha \delta_{\beta\gamma} + u_\beta \delta_{\alpha\gamma} +
u_\gamma \dab)
\label{eq:constraints}
\end{eqnarray}
where $P_{\alpha\beta}$ is the pressure tensor, $c$ is defined to be
$\dr/\dt$ and the last term of the third expression in
\eq{eq:constraints} is included to ensure Galilean invariance.

Considering a $D3Q15$ lattice and a square-gradient approximation to
the interface free energy ($\kappa (\partial_\alpha n)^2
/2$)~\cite{swift:96}, a possible choice of the coefficients
is~\cite{pooley:03}
%
\begin{eqnarray}
A_\sigma & = & \frac{w_\sigma}{c^2}\left( p_b- 
               \frac{\kappa}{2} (\partial_\alpha n)^2               
               -\kappa n \partial_{\alpha\alpha} n 
               + \nu u_\alpha \partial_\alpha n \right), \nonumber \\
B_\sigma & = & \frac{w_\sigma n}{c^2}, \quad 
  C_\sigma = -\frac{w_\sigma n}{2 c^2}, \quad 
  D_\sigma = \frac{3 w_\sigma n}{2 c^4}, \nonumber \\
G_{1\gamma\gamma} & = & \frac{1}{2 c^4} \left( \kappa(\partial_\gamma n)^2 +2 
\nu u_\gamma \partial_\gamma n \right) , \quad 
  G_{2\gamma\gamma} = 0, \nonumber \\
G_{2\gamma\delta} & = & \frac{1}{16 c^4} \left( \kappa (\partial_\gamma n)
  (\partial_\delta n) + \nu (u_\gamma \partial_\delta n + u_\delta
  \partial_\gamma n) \right)
\label{lb:eqCoeff}
\end{eqnarray}
where $w_1=1/3$, $w_2=1/24$, $\kappa$ is a parameter related to the
surface tension and $p_b$ is the pressure in the bulk which is defined
below. One can show~\cite{rowlinson:82} that the pressure tensor can
be written as
\begin{equation}
P_{\alpha\beta}=\left(p_b-\frac{\kappa}{2} (\partial_\gamma n )^2
                   -\kappa n \partial _{\gamma\gamma} n \right)
                   \delta_{\alpha\beta}
                   +\kappa (\partial_\alpha n)(\partial_\beta n).
\end{equation}


\subsection{Wetting boundary conditions}

In this paper, we will focus our attention on flat substrates normal
to the $z$ direction. The derivatives in that direction should then be
handled in such a way that the wetting properties of the substrate can
be controlled. A boundary condition can be established using the Cahn
model~\cite{cahn:77}. He proposed adding an additional surface free
energy $\Psi_c(n_s)=\phi_0-\phi_1 n_s+\cdots$ at the solid surface where
$n_s$ is the density at the surface. Neglecting the second order terms
of $\Psi_c(n)$ and minimizing $\Psi_b+\Psi_c$ (where $\Psi_b$ is the
free energy in the bulk), a boundary condition valid at $z=0$
emerges~\cite{briant:02}
\begin{equation}
\partial_z n= - \frac{\phi_1}{\kappa}.
\label{eq:cahn1}
\end{equation}
\Eq{eq:cahn1} is imposed on the substrate sites to implement the Cahn
model in the lattice Boltzmann approach. Details are given
in~\cite{briant:02}.

The Cahn model can be used to relate $\phi_1$ to $\theta$ the contact
angle defined as the angle between the tangent plane to the droplet
and the substrate. Within the Cahn model the surface tension at the
interfaces is given by~\cite{degennes:85}
\begin{eqnarray}
\sigma_{lg} & = & \int_{n_g}^{n_l} \sqrt{2\kappa W(n,T)} dn \nonumber \\
\sigma_{sl,sg} & = & \left| \int_{n_s}^{n_{g,l}} \sqrt{2\kappa W(n,T)} dn \right|+ \phi_0 -\phi_1 n_s
\label{eq:cahn2}
\end{eqnarray}
where $W(n,T)$ is the excess free energy, $\sigma_{lg}$,
$\sigma_{sg}$, $\sigma_{sl}$ are the surface tensions at the
liquid-gas, solid-gas and solid-liquid interface respectively and
$n_s$, $n_l$, $n_g$ are the densities at the substrate, of the liquid
phase and of the gas phase respectively. Young's law~\cite{young:1805}
gives a relation between the static contact angle and the surface
tensions of the three phases
\begin{equation}
\sigma_{lg} \cos \theta = \sigma_{sg} - \sigma_{sl}.
\label{eq:young}
\end{equation}

A convenient choice of bulk pressure is~\cite{briant:02}
\begin{equation}
p_b=p_c(\nu_p+1)^2(3\nu_p^2-2\nu_p+1-2\beta\tau_p)
\label{eq:pressureA}
\end{equation}
where $\nu_p=(n-n_c)/n_c$, $\tau_p=(T_c-T)/T_c$ and $p_c=1/8$,
$n_c=3.5$ and $T_c=4/7$ are the critical pressure, density and
temperature respectively and $\beta$ is a constant typically equal to
$0.1$. The excess free energy then becomes~\cite{briant:02}
\begin{equation}
W(n,T)=p_c(\nu_p^2-\beta\tau_p)^2.
\label{eq:W1}
\end{equation}

Inserting \eq{eq:W1} into relation (\ref{eq:cahn2}) and using
\eq{eq:young}, gives the relation between $\phi_1$ and $\theta$
\begin{equation}
\phi_1=2 \beta \tau_p \sqrt{2p_c \kappa} \; \mathrm{sign}(\theta -
\frac{\pi}{2}) \sqrt{\cos\frac{\alpha}{3}\left(1-\cos\frac{\alpha}{3}\right)}
\label{eq:phi1}
\end{equation}
where $\alpha=\mathrm{cos}^{-1}(\sin^2\theta)$ and the function
$\mathrm{sign}$ returns the sign of its argument.



We impose a no-slip boundary condition on the velocity. Because a
collision takes place on the boundary the usual bounce-back condition
must be extended to ensure mass conservation (see \cite{dupuis:02} for
a wider discussion). This is done by a suitable choice of the rest
field, $f_0$, to correctly balance the mass of the system.
%

\section{Spreading on a homogeneous surface}
\label{sec:validation}

\begin{figure}
\begin{center}
\begin{tabular}{ccc}
$t=0$ & $t=250$ & $t=500$ \\
\epsfig{file=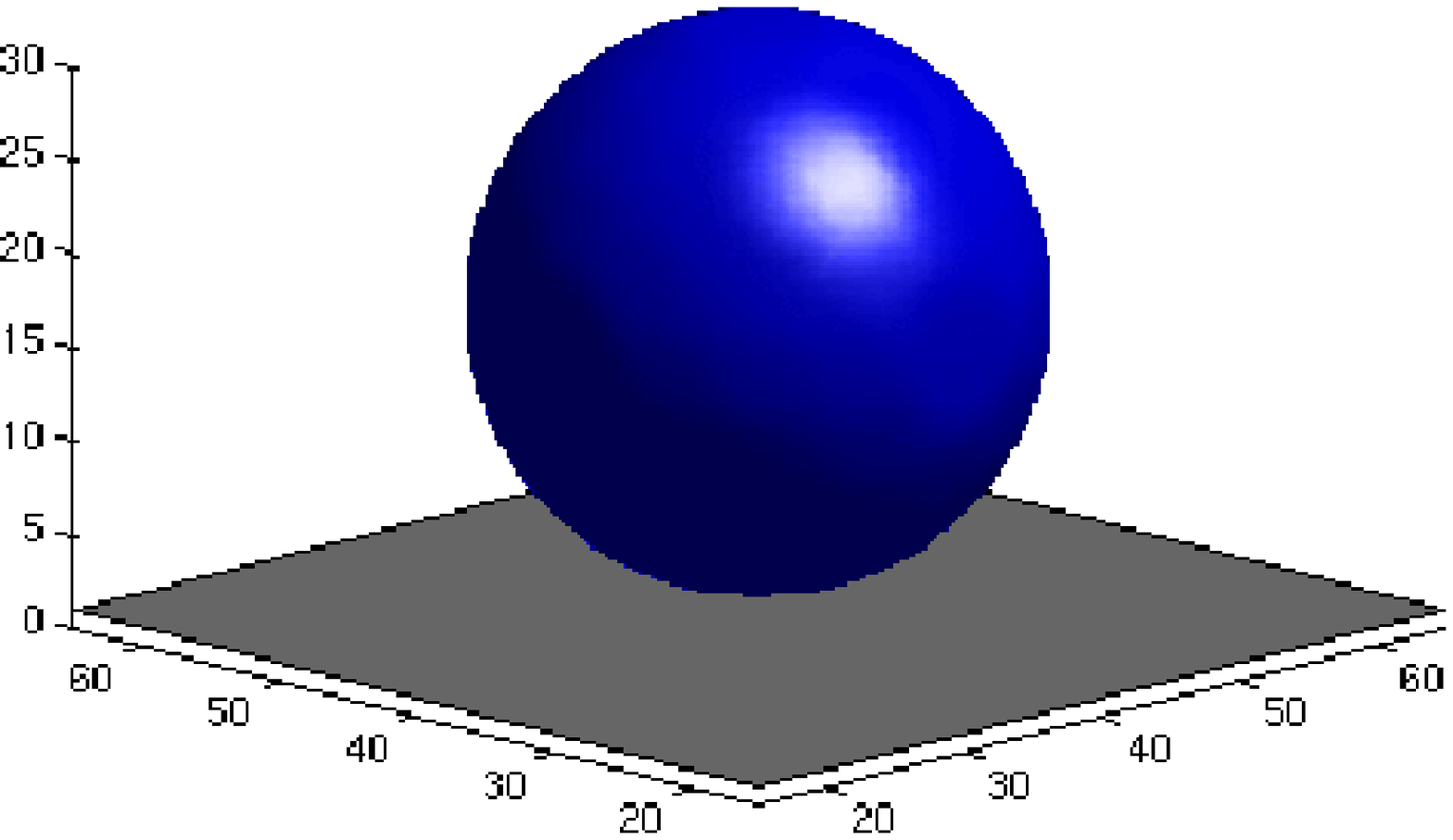,width=4cm} &
\epsfig{file=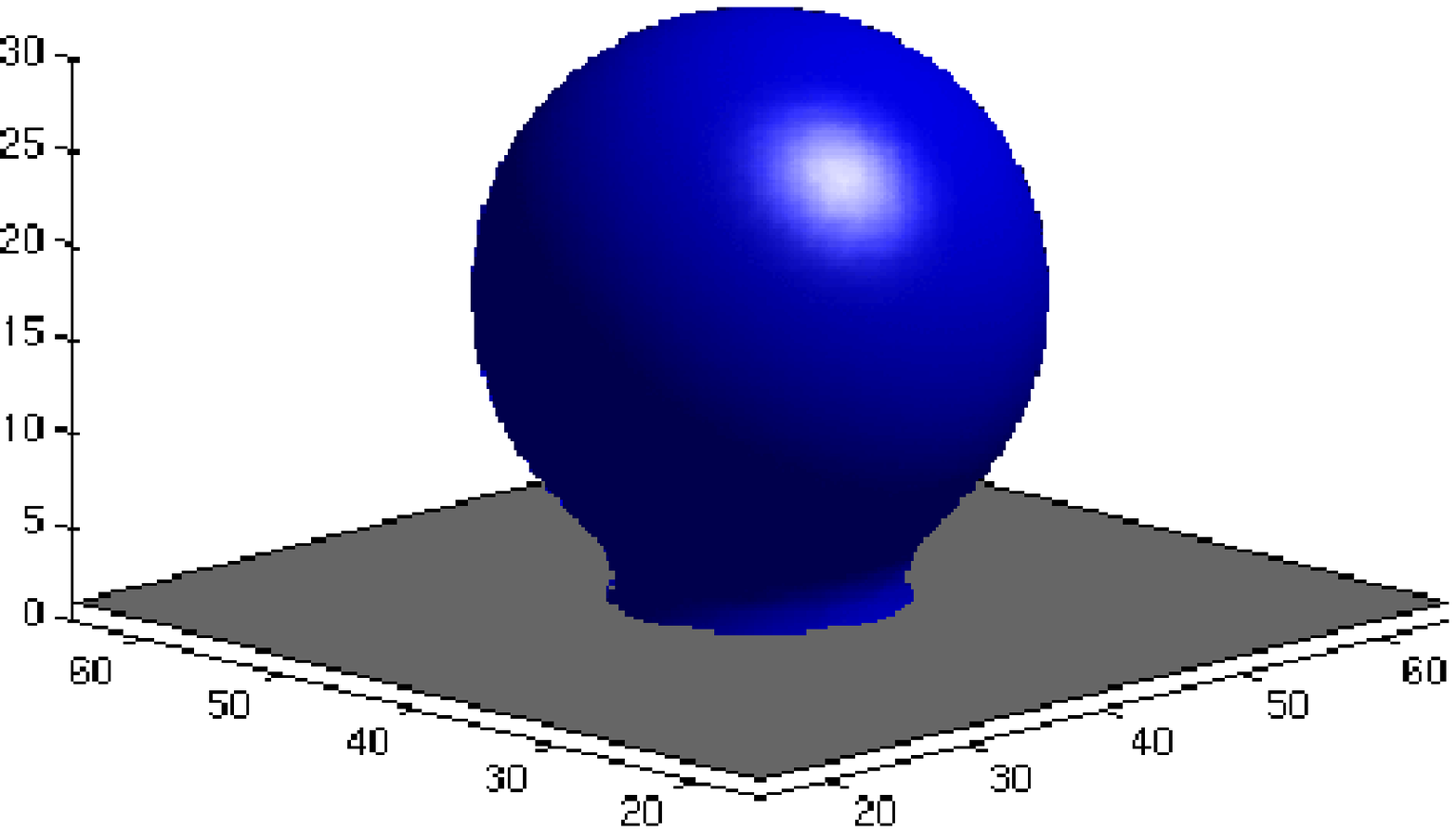,width=4cm} &
\epsfig{file=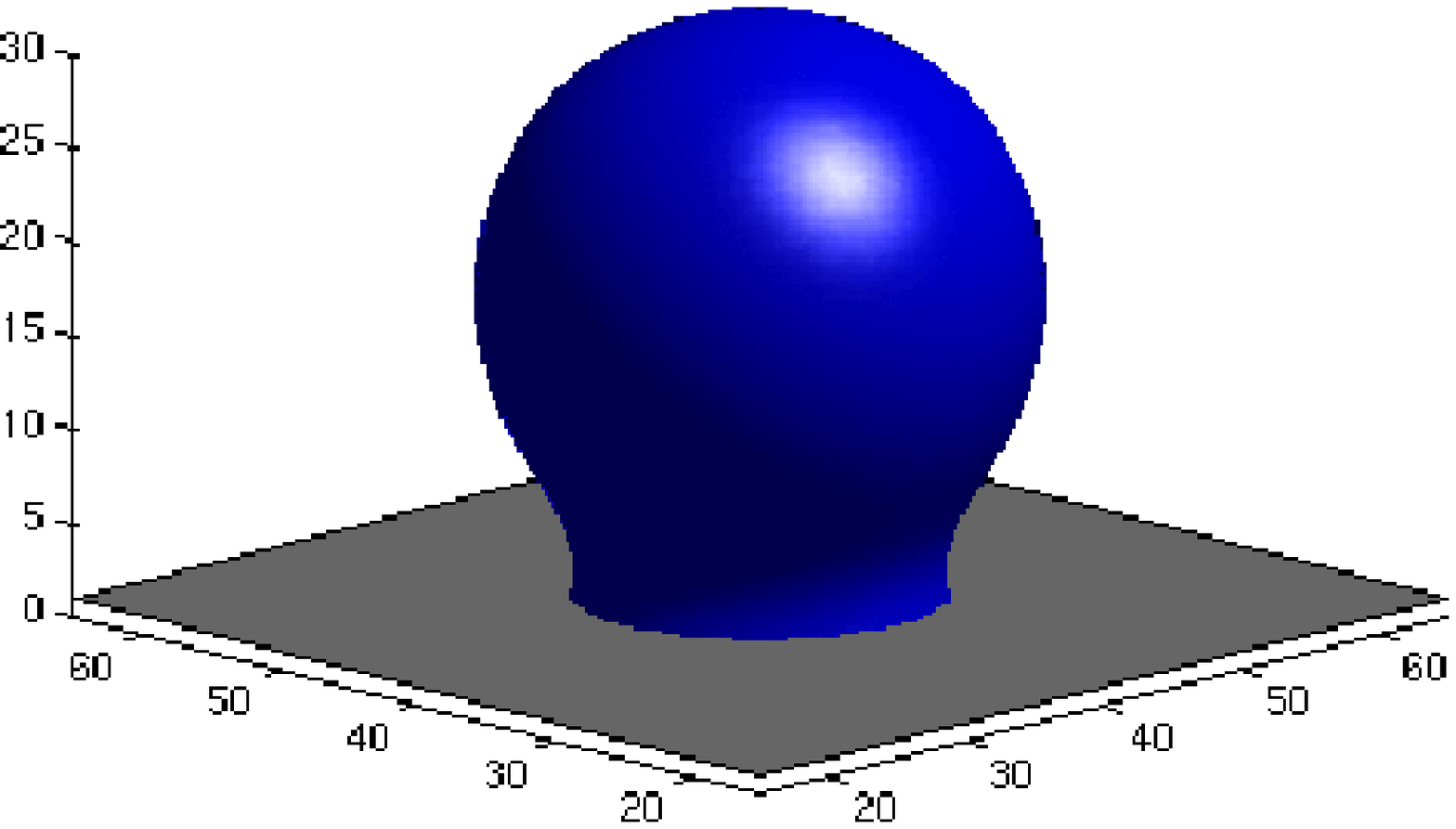,width=4cm} \\
$t=1000$ & $t=5000$ & $t=20000$ \\
\epsfig{file=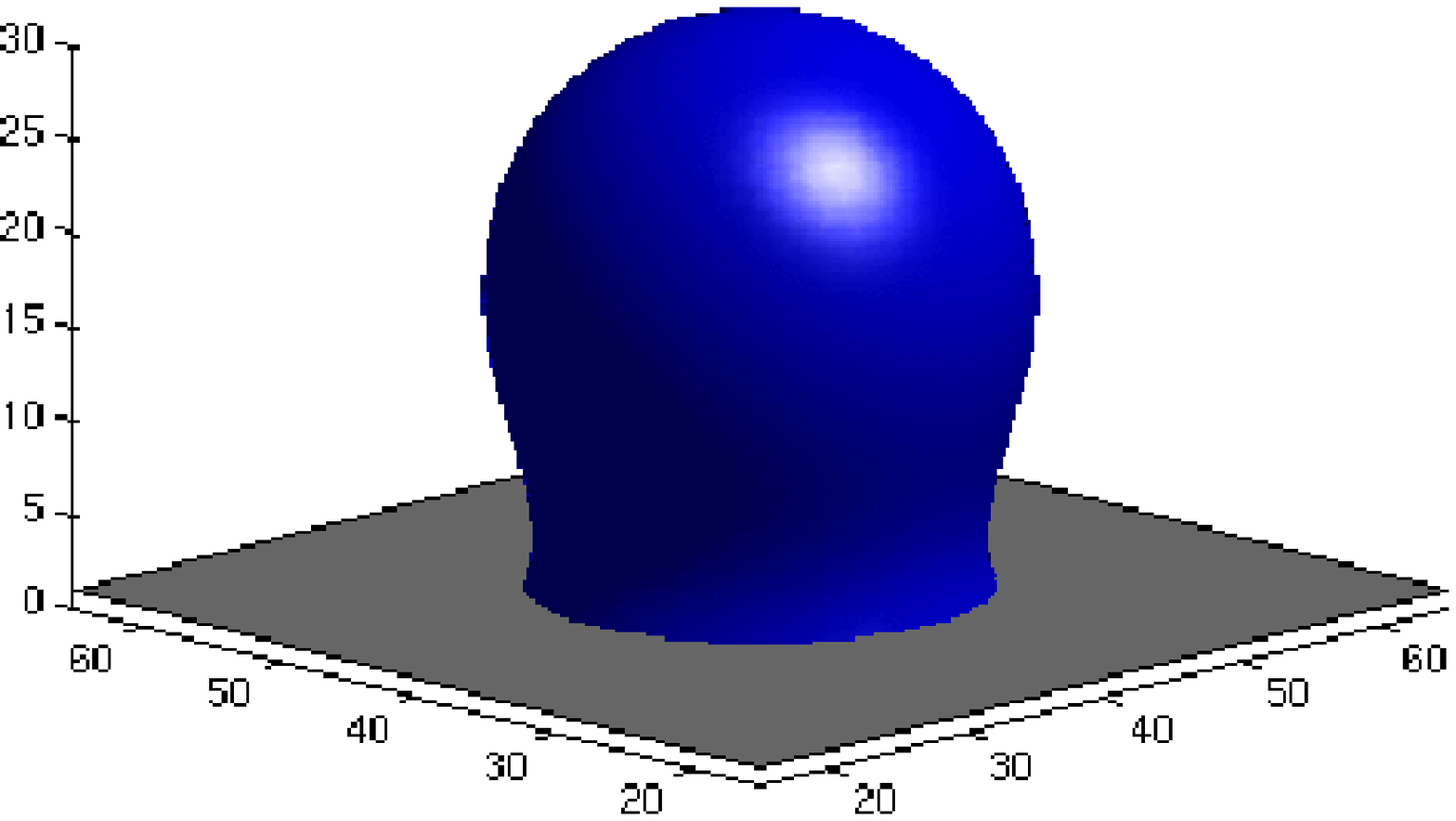,width=4cm} &
\epsfig{file=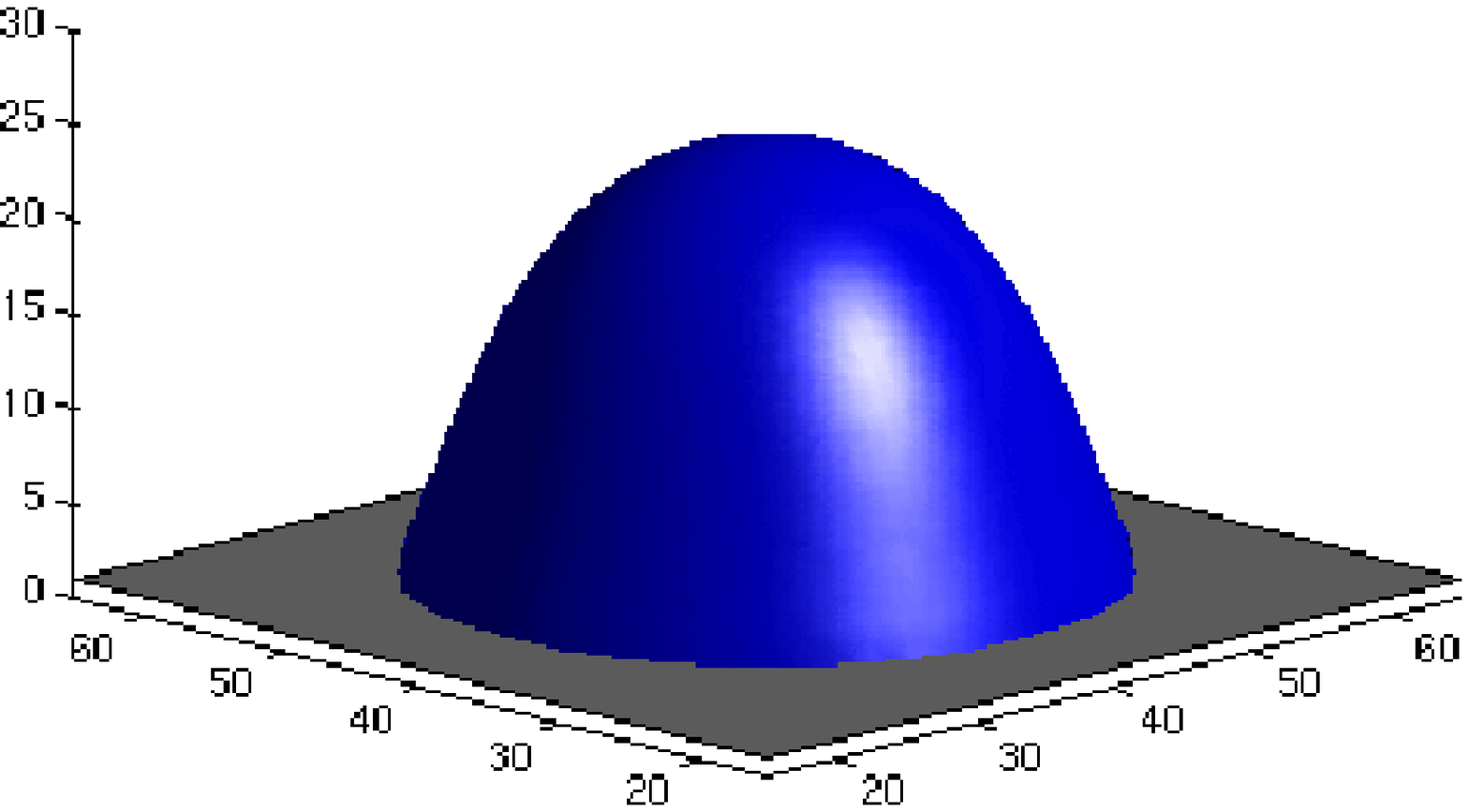,width=4cm} &
\epsfig{file=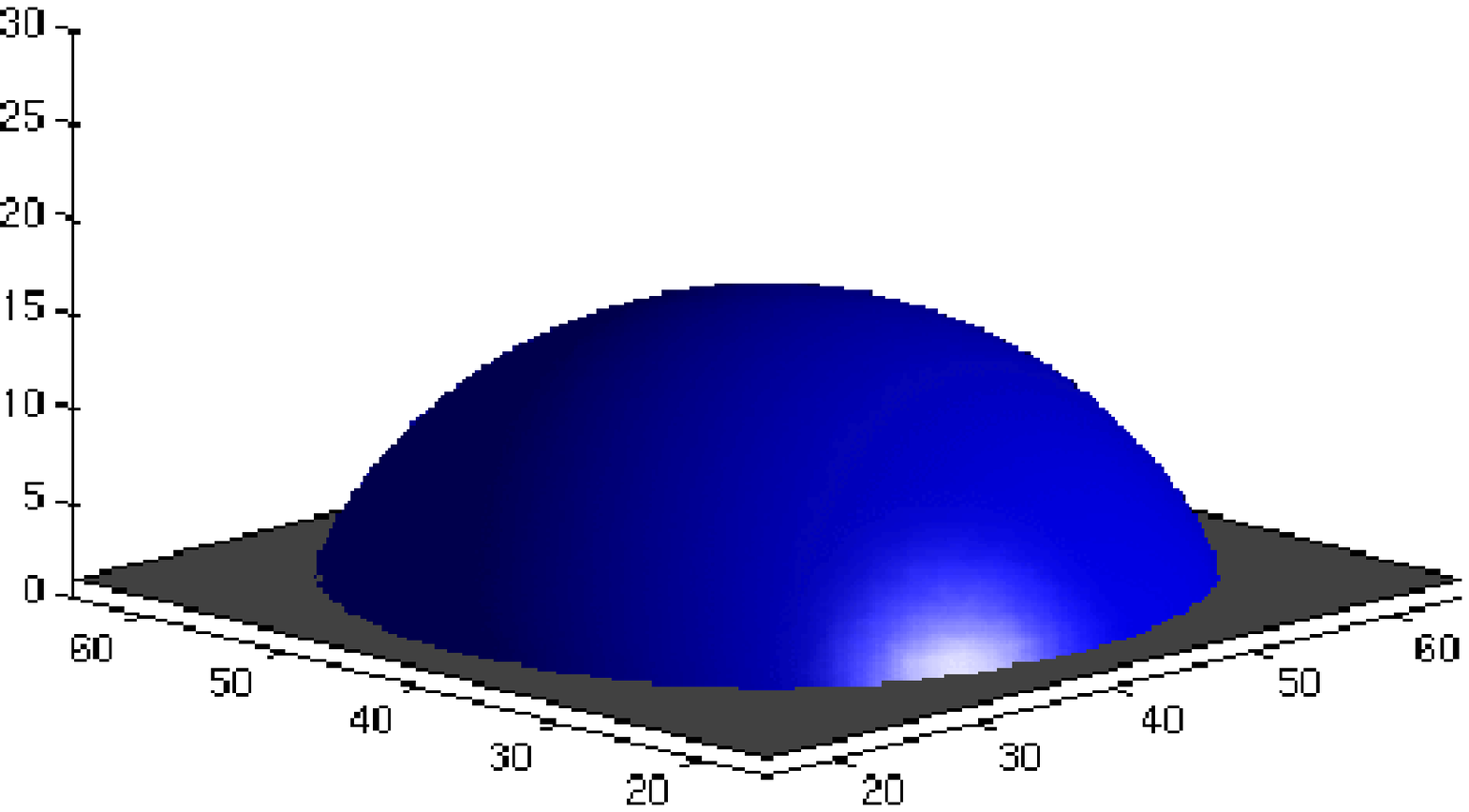,width=4cm} \\
\end{tabular}
\end{center}
\caption{Spreading of a spherical droplet of radius $R_0=16$ on a
  $80\times 80 \times 40$ lattice. The equilibrium contact angle is
  $60^\circ$.}
\label{fig:spread}
\end{figure}

We consider a $80\times 80 \times 40$ lattice on which a spherical
drop of radius $R_0=16$ just touches a flat surface at $z=0$.  Unless
otherwise specified the temperature is $T=0.4$ which leads to two
phases of density $n_l=4.128$ and $n_g=2.913$. \Fig{fig:spread} shows
how the droplet evolves in time to reach an equilibrium shape with
contact angle $60^\circ$.

\begin{figure}
\begin{center}
\epsfig{file=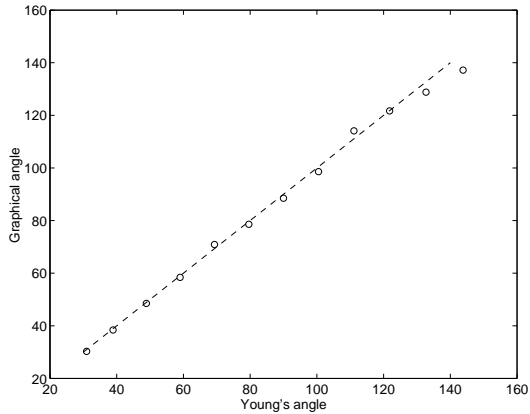,width=7cm}
\end{center}
\caption{Comparison between equilibrium contact angles $\theta^y$ and
  $\theta^g$ (defined in the text) on a $110 \times 110 \times 50$
  lattice. The input contact angles are set from $30^\circ$ to $140^\circ$
  every $10^\circ$. $\tau=1.0$ and $\kappa=0.003$. The initial droplet has
  a radius $R_0=18$. $80~000$ iterations were used to reach each
  equilibrium. The dashed line is the expected agreement.}
\label{fig:young}
\end{figure}

We first present a check on the accuracy of the equilibrium properties
of the model. \Fig{fig:young} reports a comparison between two methods
of measuring the contact angle. $\theta^y$ is the contact angle
obtained from \eq{eq:young} with the surface tensions measured at
equilibrium.  $\theta^g$ is the contact angle measured from the
profile of the simulated droplet once equilibrium is reached. The
agreement is good.  Small errors results from the difficulty of a
direct measurement of the contact angle on a discrete lattice.

The shape of the area formed by the contact of a droplet with a
homogeneous substrate is a disk. Its radius $R_c$ is a quantity which
is rather simple to measure and has consequently attracted the
attention of many scientists, see~\cite{marmur:83} and references
therein. The time evolution of $R_c$ has been found to follow a power
law $R_c=m t^{n/2}$. The exponent $n$ has been widely reported in the
literature but with no consistent result. Marmur~\cite{marmur:83} in
his review lists exponents between $0.06$ and $0.6$. The value of
$m$ appears to be related to the droplet volume.

\begin{figure}
\begin{center}
\begin{tabular}{cc}
\epsfig{file=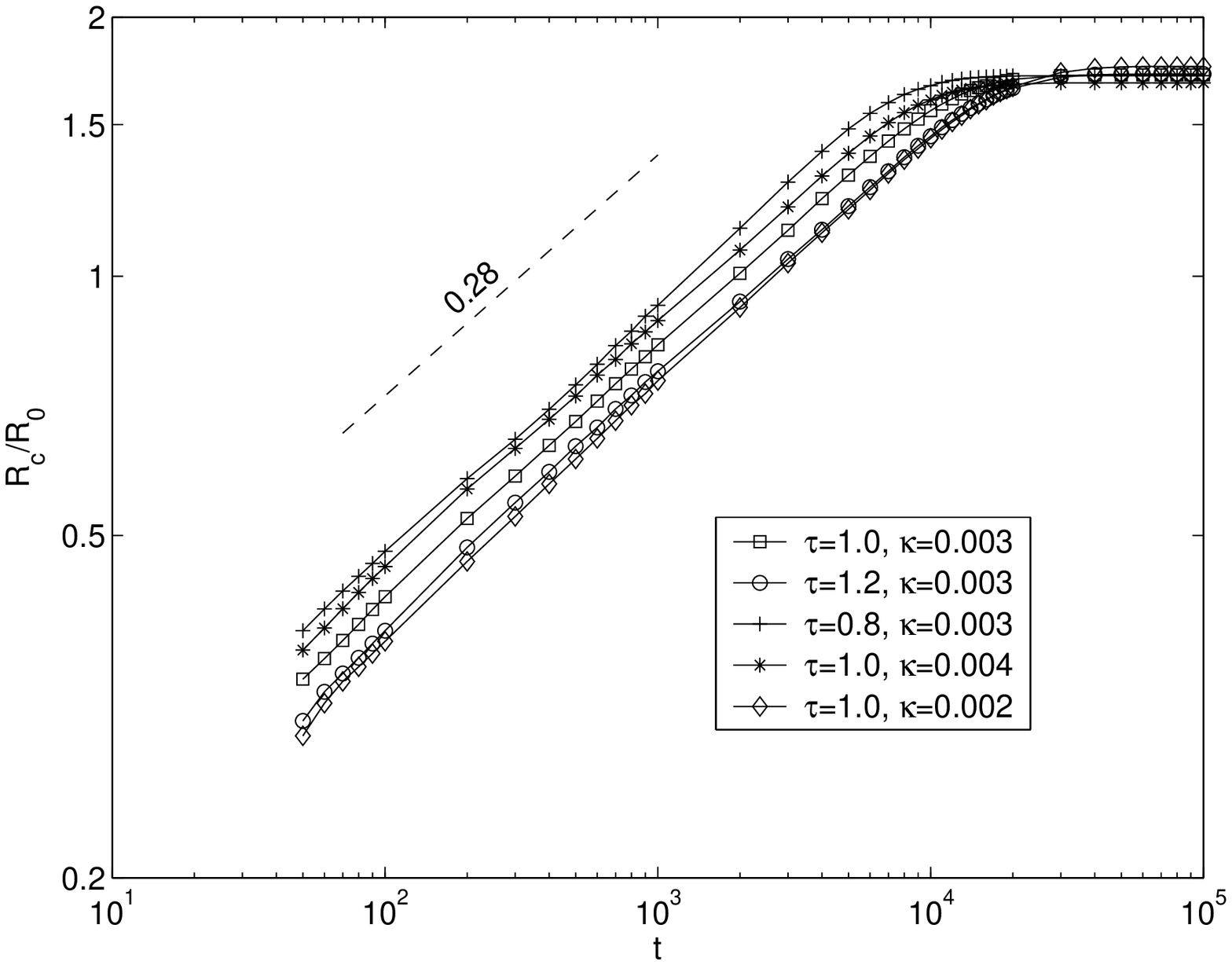,width=6cm} & \epsfig{file=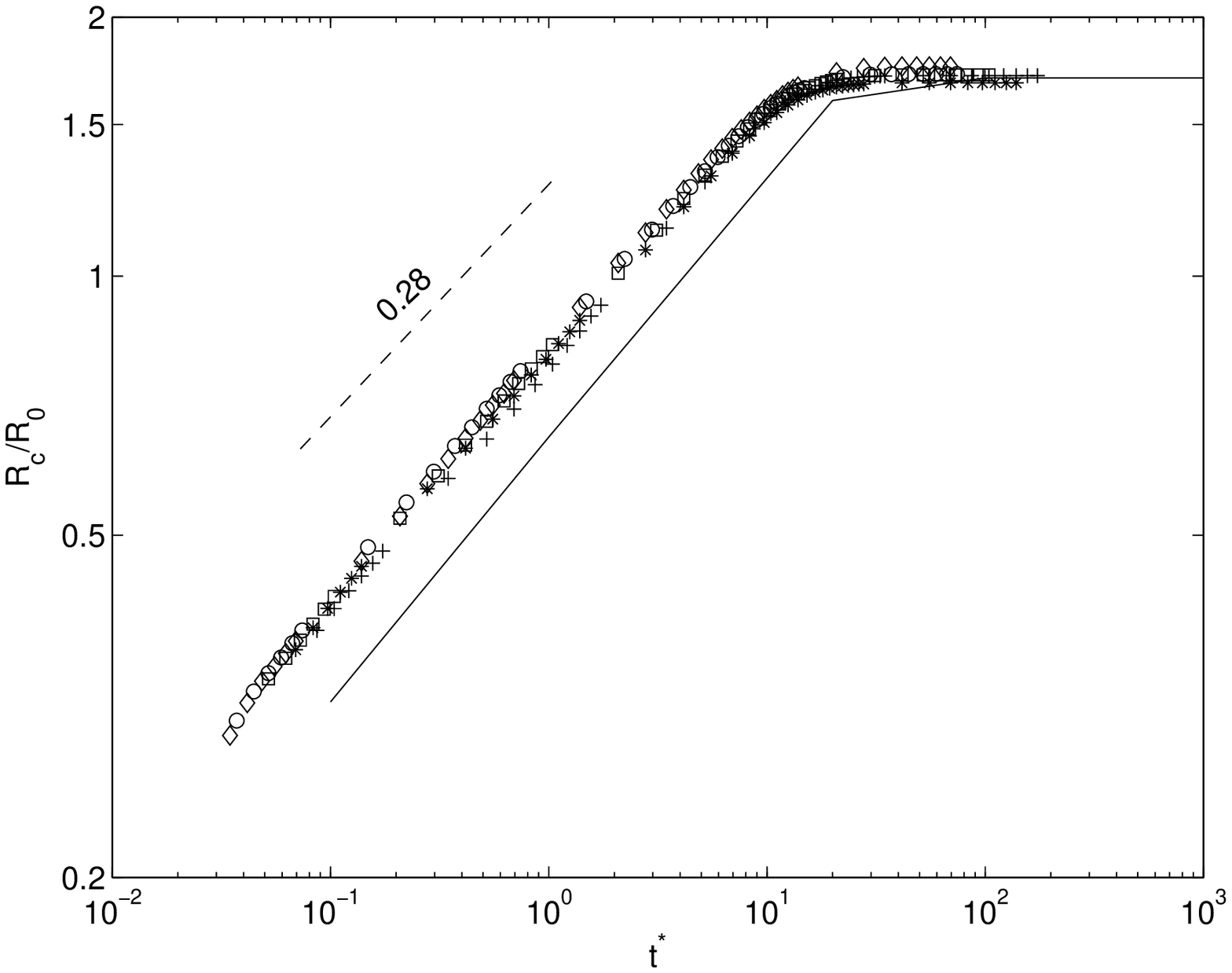,width=6cm} \\
(a) & (b) \\
\end{tabular}
\end{center}
\caption{Time evolution of the radius of the droplet base $R_c$ on 
  a $90 \times 90 \times 50$ lattice (a) as a function of time $t$ (b)
  as a function of dimensionless time $t^*$. The contact angle is set
  to $60^\circ$ and $R_0=16$. The solid line is the result of
  laboratory experiments~\cite{zosel:93}.}
\label{fig:time}
\end{figure}

\Fig{fig:time} shows the time evolution of $R_c$ for different values
of the viscosity and the surface tension. The curves correspond to a
value $n=0.56$ which is within the range reported in the literature.
The power law is independent of the surface tension and the viscosity.

Indeed if the evolution curves are plotted as a function of the
dimensionless time~\cite{zosel:93}
\begin{equation}
t \longrightarrow t^* = \frac{\sigma_{lg}}{\eta R_0} t
\end{equation}
the data collapses onto a single curve as shown in \fig{fig:time}(b).
Experimental data taken from~\cite{zosel:93} shows similar behaviour.

\section{Spreading on heterogeneous surfaces}
\label{sec:inhomo}

Almost any surface will contain physical and chemical inhomogeneities
which will affect the spreading of a mesoscopic droplet.  It has
recently become feasible to fabricate surfaces with well-defined
chemical properties on micron length scales and it is becoming
possible to perform well-controlled experiments which probe the
behaviour of mesoscopic droplets on chemically and physically
heterogeneous substrates. Thus it is particularly interesting to
develop techniques to model the effect of these surfaces on the
spreading properties of a droplet.

One of the simplest heterogeneous surfaces can be formed by
alternating stripes of materials with different wetting properties.
The static properties of droplets on such substrates have been
discussed~\cite{drelich:94,pompe:98,cassie:48}. However less attention
has been paid to the dynamics of the spreading.

\begin{figure}
\begin{center}
\begin{tabular}{m{6cm}m{0.5cm}m{5.5cm}}
\centerline{(a)} && \centerline{(b)} \\
\epsfig{file=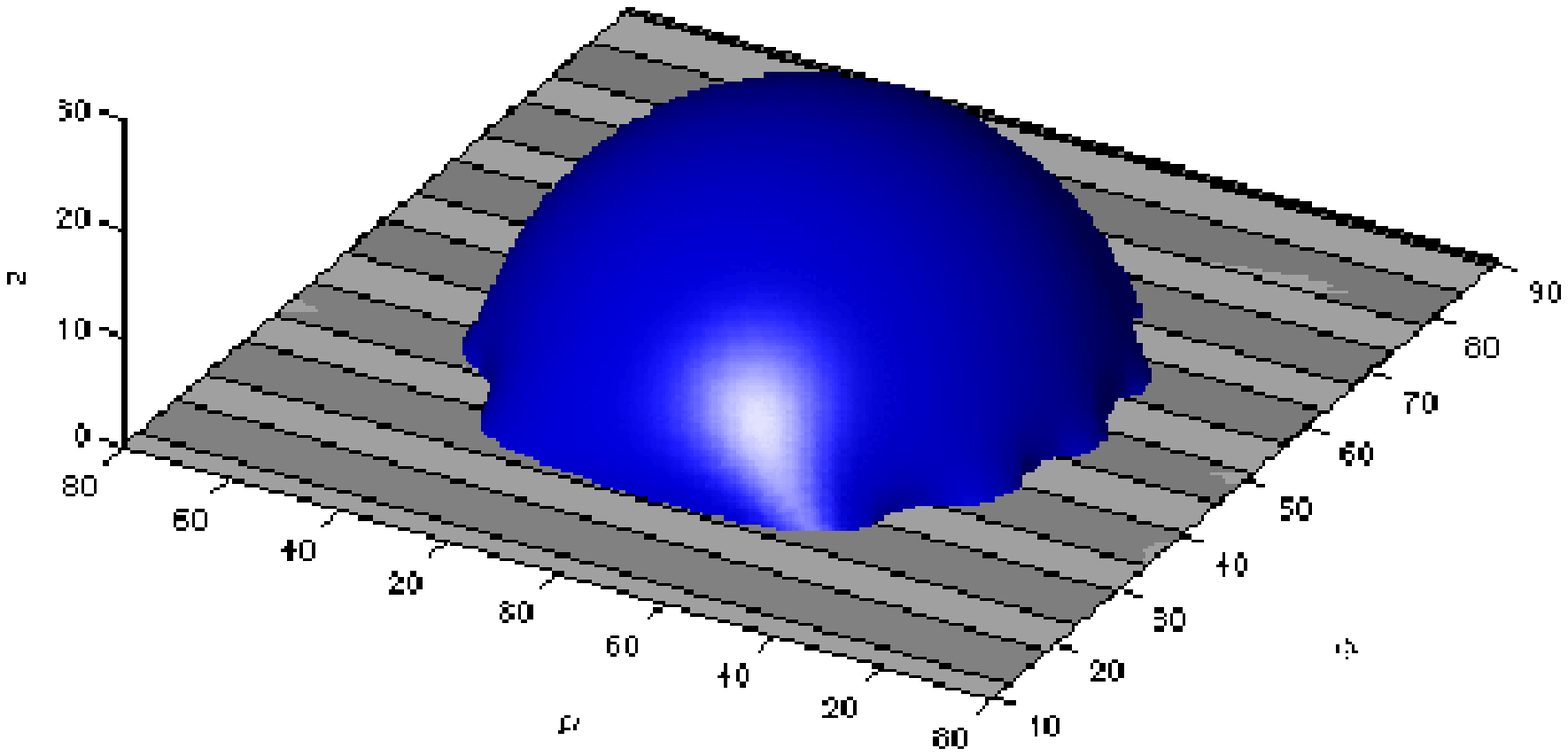,width=6cm} &&
\epsfig{file=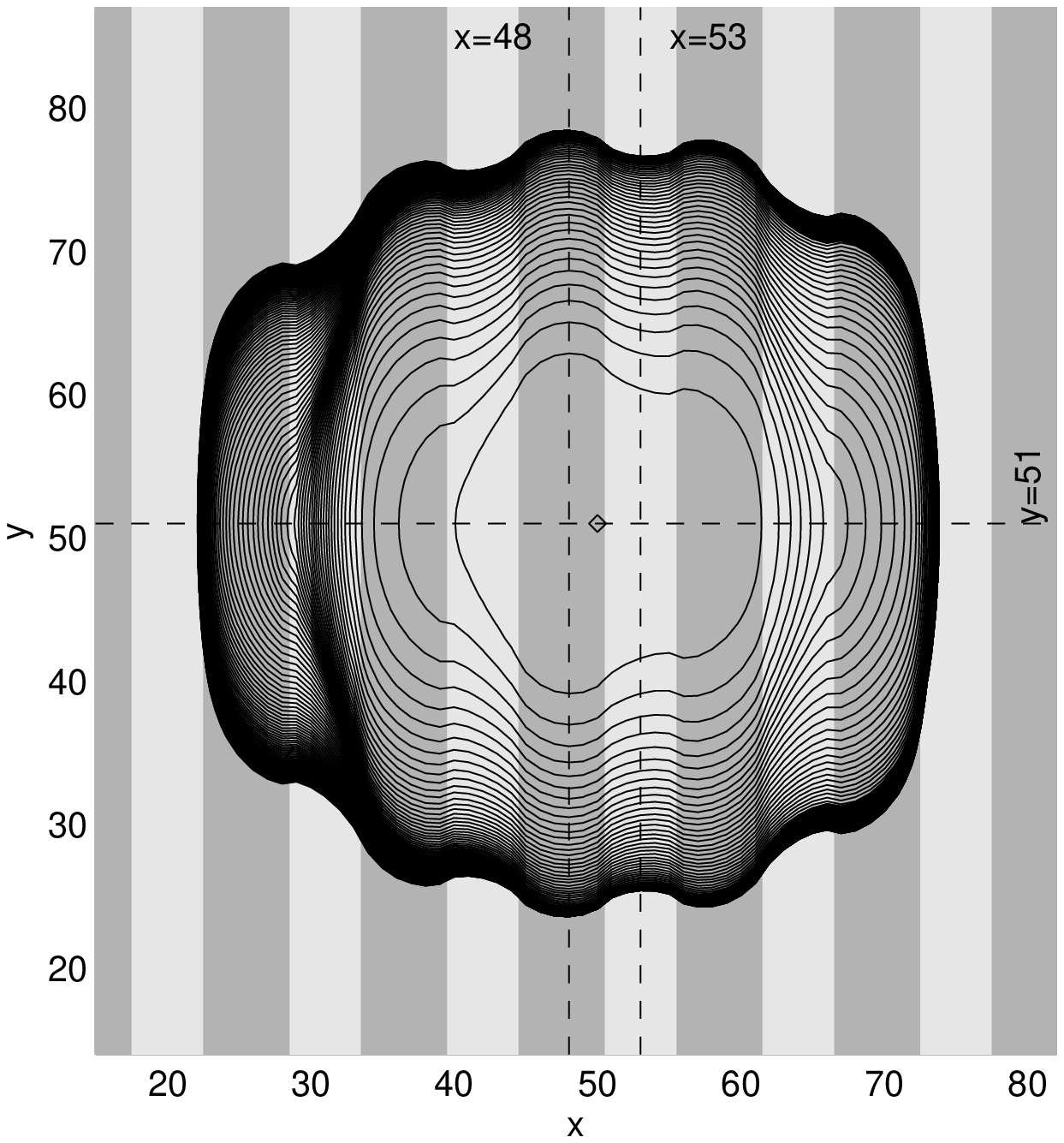,width=5.5cm}
\end{tabular}
\begin{tabular}{m{1cm}m{8cm}}
(c) & \epsfig{file=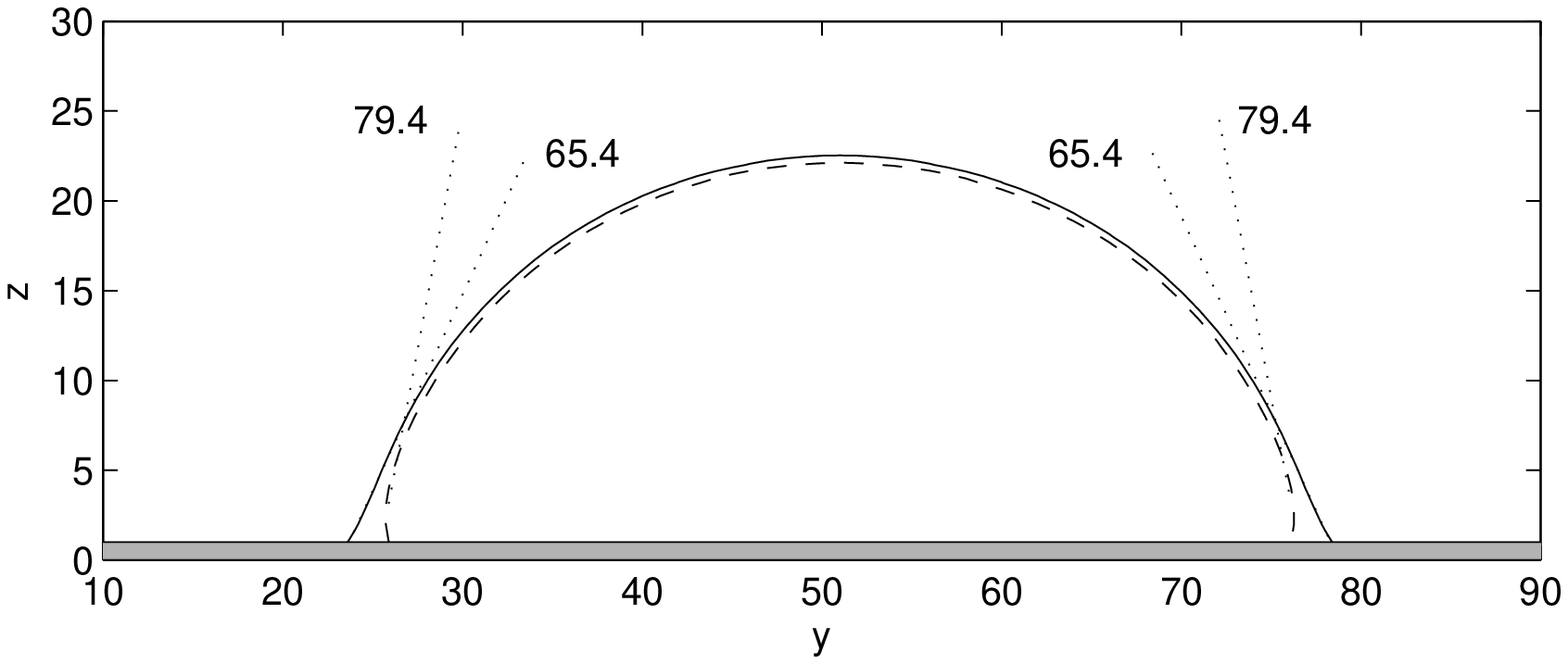,width=8cm} \\
(d) & \epsfig{file=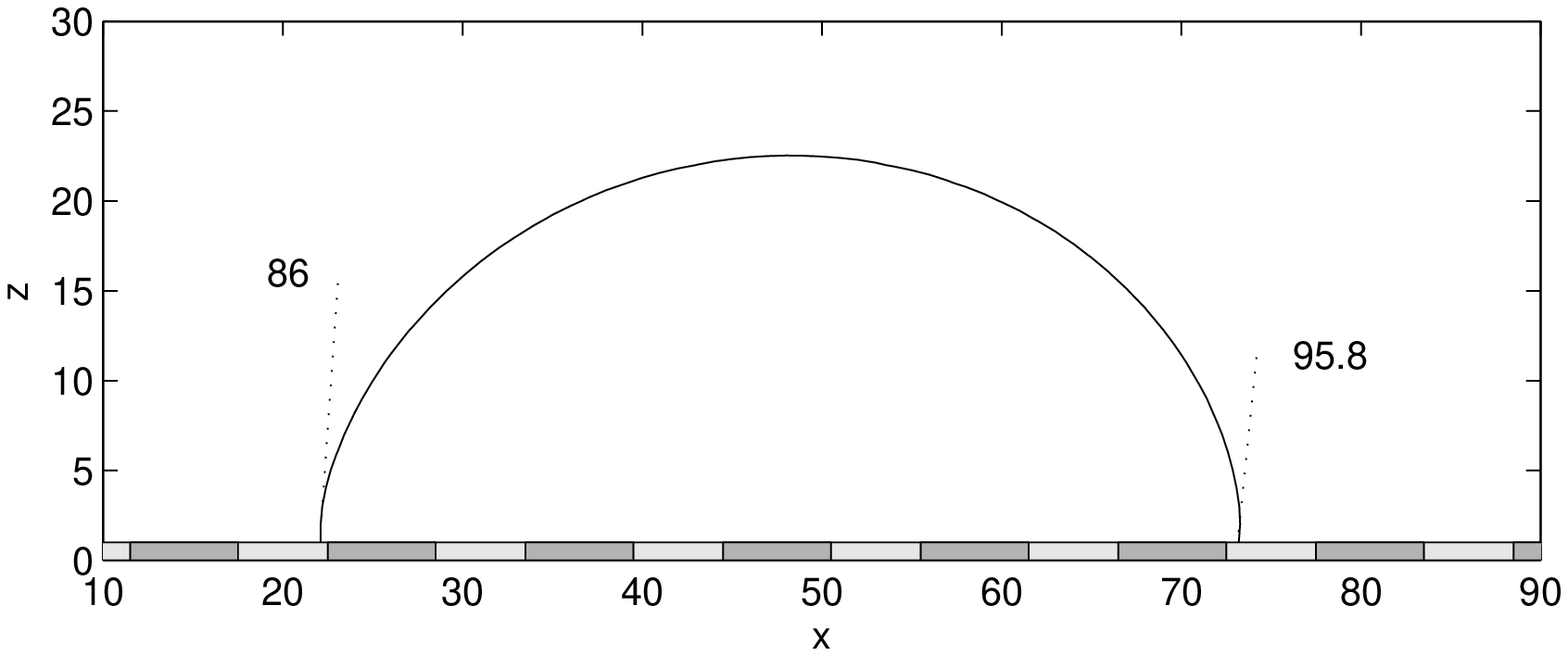,width=8cm} \\
\end{tabular}
\end{center}
\caption{Spreading of an initially spherical droplet on a heterogeneous
  surface formed by alternating hydrophilic ($\theta_{phi}=50^\circ$,
  dark grey) and hydrophobic ($\theta_{pho}=110^\circ$, light grey)
  stripes of width equal to $6$ and $5$ respectively. A $99\times
  99\times 60$ lattice and a droplet with an initial radius $R_0=19$
  are used. The droplet initially just touches the substrate at
  $x=50$, $y=51$.  $\tau=1.0$ and $\kappa=0.003$. Equilibrium is
  reached after $100~000$ iterations. (a) Three-dimensional view of
  the droplet at equilibrium. (b) Time evolution of the contact line.
  Each contour corresponds to $1000$ iterations. (c) Cross section at
  $x=48$ (solid) and $x=53$ (dashed) of the droplet profile at
  equilibrium. (d) Cross section at $y=51$ of the droplet profile at
  equilibrium.}
\label{fig:spreading}
\end{figure}

In this section we consider heterogeneous surfaces formed by
alternating hydrophilic and hydrophobic stripes. They are
characterised by widths $w_{phi}$,$w_{pho}$ and contact angles
$\theta_{phi}$, $\theta_{pho}$ respectively. \Fig{fig:spreading}
presents the behaviour of a three-dimensional droplet spreading on
such a surface with $\theta_{phi}=50^\circ$, $\theta_{pho}=110^\circ$,
$w_{phi}=6$, $w_{pho}=5$. The droplet has an initial radius $R_0=18$.

It is apparent from the figure that the behaviour of the droplet
depends on whether it is on a hydrophobic or a hydrophilic stripe. The
equilibrium shape of the contact line shown in \fig{fig:spreading}(b)
reflects the pattern of the underlying substrate which is comparable
to that found in laboratory experiments~\cite{pompe:98}.

The time evolution of the contact line is also shown in
\fig{fig:spreading}(b). Note that its velocity decreases smoothly in
the $y$-direction parallel to the stripes but not in the $x$-direction
where it moves faster on the hydrophilic than on the hydrophobic
stripes. Note also that the droplet remains symmetric about an axis
perpendicular to the stripes but that the shape becomes asymmetric
about an axis parallel to the stripes, depending upon the initial
position of the center of the droplet. 

Observation of the movement of the contact line in the $x$-direction
shows that in a hydrophilic region the contact angle tends to decrease
and the velocity of the contact line increase. When the contact line
reaches the boundary its progress is stopped and the contact angle
increases until it is large enough to cross the hydrophobic stripe.

It has been proposed that an equilibrium droplet on such a surface has
a spatially averaged contact angle following Cassie's
law~\cite{cassie:48}
\begin{equation}
\cos \theta = p_{phi}\cos \theta_{phi} + p_{pho} \cos\theta_{pho}
\label{eq:cassie}
\end{equation}
where $p_{phi}$ and $p_{pho}$ are the proportion of the substrate area
which are hydrophilic or hydrophobic respectively. However, this
relation is not universally accepted. In particular it has been argued
that there should be a dependence on the relative size of the droplet
and the surface stripes~\cite{drelich:94}.
\Fig{fig:spreading}(c)-(d) show characteristic angles for the
droplet considered here. Their average is $76.5^\circ$ which is close to
the one predicted by Cassie's law, $78.7^\circ$.

\section{Conclusion}
\label{sec:conclusion}

We have used a three-dimensional lattice Boltzmann algorithm to model
the spreading of a mesoscopic droplet. By incorporating the Cahn
theory of wetting into the simulation we obtain a way of easily tuning
the contact angle of the droplet on the substrate. This gives us the
ability to simulate spreading on both homogeneous and heterogeneous
surfaces.

The approach provides a well-controlled way of investigating the
dependence of the spreading on such properties as the droplet volume,
contact angles and the substrate geometry. Further work is in progress
to systematically determine how these parameters affect the velocity
and shapes of the spreading droplets. It would also be interesting to
investigate the effect of physical inhomogeneities on the spreading and
to consider a droplet spreading on a porous surface.  

A particular aim of the work will be to compare the results to
forthcoming experiments on substrates which have chemical patterning
on mesoscopic length scale. This will allow increased understanding of
both the experimental results and the model assumptions. For example
we assume, as is the standard practice, no-slip boundary conditions on
the velocity. These may not be appropriate on short length scales near
a contact line. Moreover the liquid-gas density difference in lattice
Boltzmann models is very small compared to real fluids and it is
important to undertake further work to assess the effect of this on
the modelling results.

\section*{Acknowledgments}

We thank D.~Bucknall, J.~Leopoldes and S.~Willkins for helpful
discussions. Supercomputing resources were provided by the Oxford
Supercomputing Centre. AD acknowledges the support of the EC IMAGE-IN
project GR1D-CT-2002-00663.

\def\biblioP{/home/wytham/dupuis/Tex/Bib/} 
\bibliographystyle{unsrt}
\bibliography{\biblioP spc,\biblioP lattice-gas,/home/cumnor/yeomans/Bib/wetting}

\begin{thebibliography}{10}

\bibitem{degennes:85}
P.G. de~Gennes.
\newblock Wetting: statics and dynamics.
\newblock {\em Review of {M}odern {P}hysics}, 57(3):827--863, 1985.

\bibitem{succi-book:01}
S.~Succi.
\newblock {\em The Lattice {B}oltzmann Equation, For Fluid Dynamics and
  Beyond}.
\newblock Oxford University Press, 2001.

\bibitem{swift:96}
M.R. Swift, E.~Orlandini, W.R. Osborn, and J.M. Yeomans.
\newblock Lattice {B}oltzmann simulations of liquid-gas and binary fluid
  systems.
\newblock {\em Phys. Rev. E}, 54:5051--5052, 1996.

\bibitem{shan:93}
X.~Shan and H.~Chen.
\newblock Lattice {B}oltzmann models for simulating flows with multiple phases
  and components.
\newblock {\em Phys. Rev. E}, 47:1815--1819, 1993.

\bibitem{he:98}
X.~He, S.~Chen, and G.D. Doolen.
\newblock A novel thermal model for the lattice {B}oltzmann method in
  incompressible limit.
\newblock {\em Journal of Computational Physics}, 146:282--300, 1998.

\bibitem{succi:89}
F.~Higuera S.~Succi, E.~Foti.
\newblock 3-dimensional flows in complex geometries with the lattice
  {B}oltzmann method.
\newblock {\em EuroPhysics Letters}, 10(5):433--438, 1989.

\bibitem{kendon:99}
V.M. Kendon, J.C. Desplat, P.~Bladon, and M.E. Cates.
\newblock 3d spinodal decomposition in the inertial regime.
\newblock {\em Physical Review Letters}, 83(3):576--579, 1999.

\bibitem{dupuis:00b}
A.~Dupuis and B.~Chopard.
\newblock Lattice gas modeling of scour formation under submarine pipelines.
\newblock {\em Journal of Computational Physics}, 178(1):161--174, 2002.

\bibitem{holdych:98}
D.~Holdych, D.Rovas, J.~Georgiadis, and R.~Buckius.
\newblock An improved hydrodynamics formulation for multiphase flow lattice
  {B}oltzmann models.
\newblock {\em Int. {J}. {M}od. {P}hys. {C}}, 9:1393--1404, 1998.

\bibitem{raiskinmaki:00}
P.~Raiskinm\"aki, A.~Koponen, J.~Merikoski, and J.~Timonen.
\newblock Spreading dynamics of three-dimensional droplets by the
  lattice-{B}oltzmann method.
\newblock {\em Computational Materials Science}, 18:7--12, 2000.

\bibitem{dupuis:02}
A.~Dupuis.
\newblock {\em From a lattice {B}oltzmann model to a parallel and reusable
  implementation of a virtual river}.
\newblock PhD thesis, University of Geneva, June 2002.
\newblock {\tt http://cui.unige.ch/spc/PhDs/aDupuisPhD/phd.html}.

\bibitem{pooley:03}
C.M. Pooley.
\newblock Private communication.
\newblock 2003.

\bibitem{rowlinson:82}
J.S. Rowlinson and B.~Widom.
\newblock {\em Molecular theory of capillarity}.
\newblock Oxford: {C}larendon, 1982.

\bibitem{cahn:77}
J.W. Cahn.
\newblock Critical point wetting.
\newblock {\em J. Chem. Phys.}, 66:3667--3672, 1977.

\bibitem{briant:02}
A.J. Briant, P.~Papatzacos, and J.M. Yeomans.
\newblock Lattice {B}oltzmann simulations of contact line motion in a
  liquid-gas system.
\newblock {\em Phil. {T}rans. {R}. {S}oc. {L}ond. {A}}, 360:485--495, 2002.

\bibitem{young:1805}
T.~Young.
\newblock An essay on the cohesion of fluids.
\newblock {\em Phil. {T}rans. {R}. {S}oc. {L}ond.}, 95:65--87, 1805.

\bibitem{marmur:83}
A.~Marmur.
\newblock Equilibrium and spreading of liquids on solid surfaces.
\newblock {\em Advances in {C}olloid and {I}nterface {S}cience}, 19:75--102,
  1983.

\bibitem{zosel:93}
A.~Zosel.
\newblock Studies of the wetting kinetics of liquid drops on solid surfaces.
\newblock {\em Colloid {P}olym. {S}ci.}, 271:680--687, 1993.

\bibitem{drelich:94}
J.~Drelich, J.~Miller, A.~Kumar, and G.~Whitesides.
\newblock Wetting characteristics of liquid drops at heterogeneous surfaces.
\newblock {\em Colloid {S}urf. {A}}, 93:1--13, 1994.

\bibitem{pompe:98}
T.~Pompe, A.~Fery, and S.~Herminghaus.
\newblock Imaging liquid structures on inhomogeneous surfaces by scanning force
  microscopy.
\newblock {\em Langmuir}, 14(10):2585--2588, 1998.

\bibitem{cassie:48}
A.~Cassie.
\newblock {\em Discuss. {F}araday {S}oc.}, 3:11, 1948.

\end{thebibliography}

\end{document}